\documentclass[journal]{IEEEtran}
\IEEEoverridecommandlockouts
\newcommand{\subparagraph}{}
\usepackage[american,fulldiode]{circuitikz}
\usepackage{epstopdf}
\usepackage{caption}

\usepackage[cmex10]{amsmath}
\usepackage{array}
\usepackage{mdwmath}
\usepackage{mdwtab}
\usepackage{caption}
\usepackage[caption=false,font=footnotesize]{subfig}
\usepackage{cite}
\usepackage{amssymb}

\newcommand{\R}{\mathbb{R}}

\usepackage{blindtext, subfig}
\usepackage{multicol}
\usepackage{amssymb}

\mathcode"65"65

\begin{document}

\title{Embedding AC Power Flow with Voltage Control in the Complex Plane : The Case of Analytic Continuation via Pad\'{e} Approximants}



\author{\IEEEauthorblockN{Sina S. Baghsorkhi\IEEEauthorrefmark{1}, and Sergey P. Suetin\IEEEauthorrefmark{2}\\}
\IEEEauthorblockA{\IEEEauthorrefmark{1}Department of Electrical Engineering and Computer Science, University of Michigan, Ann Arbor, USA\\
Email: sinasb@umich.edu\\}
\IEEEauthorblockA{\IEEEauthorrefmark{2}Steklov Mathematical Institute of the Russian Academy of Sciences, Moscow, Russia\\
Email: suetin@mi.ras.ru}\vspace{-2mm}}%

\maketitle
\pagenumbering{gobble}

\begin{abstract}
This paper proposes a method to embed the AC power flow problem with voltage magnitude constraints in the complex plane. Modeling the action of network controllers that regulate the magnitude of voltage phasors is a challenging task in the complex plane as it has to preserve the framework of holomorphicity for obtention of these complex variables with fixed magnitude. Hence this paper presents a significant step in the development of the idea of Holomorphic Embedding Load Flow Method (HELM) \cite{Trias}, introduced in 2012, that exploits the theory of analytic continuation, especially the monodromy theorem \cite{Markushevich} for resolving issues that have plagued conventional numerical methods for decades. This paper also illustrates the indispensable role of Pad\'{e} approximants for analytic continuation of complex functions, expressed as power series, beyond the boundary of convergence of the series. Later the paper demonstrates the superiority of the proposed method over the well-established Newton-Raphson as well as the recently developed semidefinite and moment relaxation of power flow problems.
\end{abstract}

\begin{IEEEkeywords}
AC power flow, voltage control, power flow feasibility, voltage stability, holomorphic functions, monodromy, analytic continuation, continued fractions, Pad\'{e} approximants, Stahl's compact set. \vspace{-1mm}
\end{IEEEkeywords}

\section{Introduction}

Power flow, the most fundamental concept in power system engineering, is at the heart of studies ranging from daily operation to long-term planning of electricity networks. AC power flow problem is a system of nonlinear algebraic equations that mathematically models the steady-state relations between the phasor representation of parameters and unknown states in an AC circuit. The parameters typically consist of power generated and consumed by source and sink nodes and the electrical properties, i.e. the impedance, of lines that connect these nodes. The unknown states are primarily voltage phasors but could also include continuous or discrete variables associated with network controllers, e.g. FACTS devices and tap-changing or phase-shifting transformers. The accurate and reliable determination of these states is imperative for control and thus for efficient and stable operation of the network. In certain studies it is equally vital to determine for which parameter values the power flow problem becomes infeasible as this condition is intimately linked to saddle-node bifurcation and the voltage collapse phenomenon~\cite{Venikov, Sauer, VC, SB}. The distance in the parameter space to power flow infeasibility can serve as a margin of voltage stability~\cite{VC}. This is certainly one of the most theoretical areas in electrical engineering. As conventional power systems undergo a fundamental transformation by large-scale highly-variable wind and solar generation, distributed across the network, this field can experience a resurgence~\cite{SB}. Given the inherent limitations of traditional methods, deeper understanding of these complicated phenomena requires new theoretical approaches rooted in complex analysis and algebraic geometry.

The basis of power flow is Kirchhoff's current law which states that for every node $i$ in $\mathcal{N}$, the set of all nodes, $I_i$, the net current flowing out of that node, is related to its voltage $V_i$ and those of its adjacent nodes $V_k$ in the following way:\vspace{-3mm}

\begin{align}\label{}
I_i=\sum_{k\in \mathcal{N}(i)}I_{ik}= \sum_{k\in \mathcal{N}(i)} \frac{V_i-V_k}{Z_{ik}} =\sum_{k\in \mathcal{N}[i]} V_kY_{ik}
\end{align}

$\mathcal{N}(i)$ and $\mathcal{N}[i]$ are the open and closed neighborhoods of node $i$. $I_{ik}$ is the current flow through the line connecting node $i$ and $k$ and $Z_{ik}=R_{ik}+jX_{ik}$ is the impedance of that line which is used to construct the diagonal and off-diagonal elements of the admittance matrix as,\vspace{-3mm}

\begin{align}\label{}
\displaystyle Y_{ii}=\sum_{k\in \mathcal{N}(i)} \frac{1}{Z_{ik}}, \hspace{2mm} Y_{ik}=-\frac{1}{Z_{ik}}
\end{align}

Since complex power is $S_i=P_i+jQ_i=V_iI_i^*$, the power flow problem in its complex form can be expressed as,
\begin{align}\label{}
&\displaystyle S^*_i=\sum_{k\in \mathcal{N}[i]} V^*_iV_k Y_{ik} & \forall i\in  \mathcal{N}-\{r\} \label{complex}
\end{align}

Here $r$ is the voltage reference node with $|V_r|=\text{constant}$ and $\arg(V_r)=0$. It also serves as the slack node meaning that $S_r$ is a free parameter that accounts for the mismatch of complex power and its losses throughout the network.

The numerical methods, developed historically to solve this problem, take the polynomial system of (\ref{complex}) out of its complex form by reformulating it either in rectangular form as in (\ref{rect}) or in polar form as in (\ref{polar}) where $\displaystyle V_i = e_i + j f_i= |V_i|\exp(j\theta_i)$ and $Y_{ik} =G_{ik} + j B_{ik}$. These techniques, all based on Newton's method or its variants, iteratively linearize and approximate $P_i$ and $Q_i$ in (\ref{rect}) or (\ref{polar}), starting from an initial guess.\vspace{-2mm}

\begin{subequations}\label{rect}
\begin{align}
\label{opf_Pbalance}
\hspace{-3mm}P_i = e_i\hspace{-1mm} \sum_{k\in \mathcal{N}[i]}\hspace{-2mm} \left( G_{ik} e_k - B_{ik} f_k \right) + f_i\hspace{-1mm} \sum_{k\in \mathcal{N}[i]}\hspace{-2mm} \left( B_{ik}e_k + G_{ik}f_k \right) \\
\label{opf_Qbalance}
\hspace{-3mm}Q_i = f_i\hspace{-1mm} \sum_{k\in \mathcal{N}[i]}\hspace{-2mm} \left( G_{ik} e_k - B_{ik} f_k \right) - e_i\hspace{-1mm} \sum_{k\in \mathcal{N}[i]}\hspace{-2mm} \left( B_{ik}e_k + G_{ik}f_k \right)
\end{align}
\end{subequations}

\begin{subequations}\label{polar}
\begin{align}
\label{opf_Pbalance}
P_i = |V_i| \sum_{k\in \mathcal{N}[i]} \left |V_k|(G_{ik}\cos\theta_{ik}+B_{ik}\sin\theta_{ik} \right)\\
\label{opf_Qbalance}
Q_i = |V_i| \sum_{k\in \mathcal{N}[i]} \left |V_k|(G_{ik}\sin\theta_{ik}-B_{ik}\cos\theta_{ik} \right)
\end{align}
\end{subequations}



There are two inherent shortcomings in such methods that can arise near the feasibility boundary of~\eqref{rect} or \eqref{polar} characterized by the saddle-node bifurcation manifold in their parameter spaces. Physically, proximity to the feasibility boundary corresponds to a network operating close to its loadability limit such as periods of peak electricity demand. The first issue is the poor convergence of the tangent-based search methods, likely due to the increased curvature of the hypersurfaces in~\eqref{rect} or~\eqref{polar}. The second issue near the feasibility boundary where two or more algebraic branches coalesce at a branch point is convergence to solutions that lie on other algebraic branches. Although dependent on the dynamical model of the physical system, these solutions in power systems typically signify unstable~\cite{VC} or low voltage~\cite{Sauer2} operating points. Most of these operating points cannot be physically realized and are thus false solutions. The region of initial guesses in Newton's method that converge to a particular solution has a fractal boundary. The multiple fractal domains of convergence are pressed together near the bifurcation manifold which explains erratic behavior of such methods in finding the desirable, i.e. stable/high voltage, solution even with seemingly reasonable initial guesses~\cite{Thorp}.

Recently a semidefinite relaxation of rectangular power flow in~\eqref{rect} has been reformulated as a special case of optimal power flow where the objective function of the semidefinite programming (SDP) is minimizing active power loss~\cite{DM}. This addresses the convergence failure of iterative methods but has its own serious drawbacks. First, the relaxation may not be tight and yield a high rank matrix where it is impossible to directly recover any solution to the original power flow, let alone the desirable one. Second, if the solution of the relaxed problem is high rank, nothing can be concluded on the feasibility of the power flow in the same vein as non-convergence of iterative methods cannot rule out the existence of solutions. Third, the suggested heuristic, i.e. active power loss minimization, does not always find the stable/high voltage solution branch. The first two problems can be remedied, at least in theory, by obtaining higher-order and thus tighter relaxations of~\eqref{rect}. The computation cost, however,  explodes with the order of relaxation and the number of variables. Reference~\cite{Lasserre} discusses the theoretical underpinning of this approach in the context of the generalized moment problem and highlights its connection to {\em real algebraic geometry} which, we see as an obstacle to distinguishing the desirable solution branch for algebraic problems of a complex analytic nature.

Among the above issues, the challenge of finding the solution on the desirable branch, more than anything else, underlines the significance of embedding the power flow problem in the complex plane where the extraordinary potentials of analytic continuation theory for multi-valued complex functions can be tapped. This is pioneered by the idea of holomorphic embedding load flow (HELM) which builds on the fact that under no load/no generation condition ($S_i=0 \hspace{2mm} \forall i\in  \mathcal{N}$), the network has a trivial non-zero solution for voltage phasors. This corresponds to all currents $I_{ik}$ being zero and reference voltage $V_r$ propagated across the network and defines the germ of the stable/high voltage branch. Analytic continuation of this germ is guaranteed by monodromy theorem to yield the desirable solution all the way to the closest bifurcation in the parameter space of~\eqref{complex} where there is a non-trivial monodromy and the physically meaningful solution ceases to exist.

Although the idea of HELM has aroused significant interest in the power system community, it yet has to prove its superiority over conventional methods. In this paper we demonstrate how the magnitude of complex variables can be held fixed while preserving the framework of holomorphicity. This is an important step in the embedding of power flow as it models the action of network controllers in the complex plane. We also show the indispensable role of Pad\'{e} approximants especially for the case of voltage controllers that fix the magnitude but not the argument of voltage phasors. Throughout the paper we refer to this method as PA to highlight the central role of rational approximation of functions of a complex variable for recovering the power flow solution. With this abbreviation we also want to emphasize the critical direction of research for further development of this method. In Section II we succinctly review the main ideas of HELM as presented in the original paper~\cite{Trias}, i.e. for the PQ buses, introduce the concept of rational approximation of analytic functions in relation to power series and continued fractions and explore the zero-pole structure of Pad\'{e} approximants for a 3-bus example. In Section III we introduce the mathematical {\em static} model of the most prevalent controller in the network, the automatic voltage regulator (AVR) of the generator. We demonstrate through modification of the previous 3-bus network, this time with a generator (PV bus), how the approximation of functions of a single complex variable is essential for analytic continuation of the voltage phasors with fixed magnitude. We interpret the zero-pole structure of Pad\'{e} approximants and its transformation as the solution reaches the feasibility boundary and explain the significance of the zero-pole distribution of the Pad\'{e} approximants in terms of voltage stability margin at a given operating point.
In section IV we demonstrate the superiority of the PA over conventional and recently developed methods of solving the power flow problem.  We introduce a 7-bus network where Newton-Raphson either fails to converge or converges to an unstable/low voltage solution as the active power output of a given generator changes. We also show that in this network first-order semidefinite relaxation only obtains the solution in a small subset of the stable solution branch and the second-order (moment) relaxation obtains the false solution branches. The numerical results of these methods are contrasted with that of the PA method which consistently obtains the stable/high voltage solution whenever it exists and declares the non-existence of a physically meaningful solution beyond the closest saddle-node bifurcation. In this network the zero-pole distribution of the Pad\'{e} approximants depicts the analytic structure of the voltage phasors and confirms the general pattern of voltage stability margin observed in section III.  In section V we summarize the key contributions of the paper and outline the ongoing work.

\section{Embedding the System of Equations in the Complex Plane}

Consider the following parametrization of~\eqref{complex} in terms of $s\in \mathbb{C}$ with $V^*_{i}$ replaced with independent variables $W_i$,\vspace{-1mm}
\begin{subequations}\label{polynomial}
\begin{align}
\label{poly1}
&\displaystyle sS^*_i=\sum_{k\in \mathcal{N}[i]}W_i V_k Y_{ik} &\forall i\in  \mathcal{N}-\{r\}\\
\label{poly2}
&\displaystyle sS_i=\sum_{k\in \mathcal{N}[i]} V_i W_k Y^*_{ik} & \forall i\in  \mathcal{N}-\{r\}
\end{align}
\end{subequations}

\vspace{-1mm}From a geometric point of view, the $2n$ equations of~\eqref{polynomial} define {\em generically} an affine algebraic curve in $(s,V_1,V_2,...,W_n)$. It follows from the Kirchhoff's current law and the existence of the voltage reference node (with $V_r$ appearing in~\eqref{polynomial} as a parameter) that the polynomials on the right side of~\eqref{poly1}-\eqref{poly2} (i.e. $\sum_{k\in \mathcal{N}[i]}W_i V_k Y_{ik}$ and $\sum_{k\in \mathcal{N}[i]} V_i W_k Y^*_{ik}$ $\forall i\in  \mathcal{N}-\{r\}$) are algebraically independent. To establish this algebraic independence in relation to the reference node requires rigorous analysis, a task which lies outside the scope of this paper. Taking the algebraic independence of these polynomials for granted, degenerate cases where the equations of~\eqref{polynomial} define not an algebraic curve but a higher-dimension algebraic variety can only arise when the power flow problem is ill-defined as in the case of networks with disconnected graphs. This is in line with the physical intuition that in the absence of a reference voltage, voltages are floating and a given $V_i$ can assume any value in $\mathbb{C}$. The equations of~\eqref{polynomial} generate an ideal and give a starting basis for finding the corresponding reduced Groebner basis~\cite{Cox}. For any lexicographic order, such as $...>V_i>s$, this gives a last basis element which is a bivariate polynomial $f_i(V_i,s)$ for well-defined problems. $f_i(V_i,s)=0$ can be solved for an algebraic (multi-valued) function $V_i = V_i(s)$ which has holomorphic branches where $\partial{f_i(V_i,s)}/\partial{V_i}\neq 0$. By permuting the order, we arrive at $2n$ algebraic functions $V_i = V_i(s)$, $W_i = W_i(s)$, $i = 1, . . . , n$, giving an algebraic parametrization of the curve defined by~\eqref{polynomial}. A branch point of this curve will be where any of the components $V_i(s)$ or $W_i(s)$ has a branch point, i.e., $f_i(V_i,s)=0$ and $\partial{f_i(V_i,s)}/\partial{V_i}=0$ (and similarly for the $W_i$). The branch point closest to the point $s_0$ at which the Taylor series expansion of any single-valued branch is developed, determines the radius of convergence of the series. Branch points play a critical role in the analytic continuation of these solutions and the PA method. This analysis also extends to the case of voltage control where we introduce new variables $S_i(s)$ and $\overline{V}_i(s)$.

Now that $V_i(s)$ is analytic in $s$,  $(V_i(s^*))^*$ is also analytic in $s$ and identical to the conjugate of $V_i(s)$ on the real axis. Hence solving~\eqref{complex} is equivalent to analytic continuation of the solution of the following system from $s=0$ to $s=1$,\vspace{-3mm}

\begin{align}
\label{p}
&\displaystyle \frac{sS_i(s)}{(V_i(s^*))^*}=\hspace{-2mm}\sum_{k\in \mathcal{N}[i]} V_k(s) Y_{ik} \hspace{20mm} \forall i\in\mathcal{N}-\{r\}
\end{align}

By defining $V_i(s)=\sum_{n=0}^{\infty} c_n^{[i]}s^n$, $1/V_i(s)=\sum_{n=0}^{\infty} d_n^{[i]}s^n$ and $1/ (V_i(s^*))^*=\sum_{n=0}^{\infty} {d^*_n}^{[i]}s^n$, this system is adequately described by the following set of power series relations,\vspace{-3mm}

\begin{subequations}\label{HELM1}
\begin{align}
\label{main}
&sS^*_i\sum_{n=0}^{\infty} {d^*_n}^{[i]}s^n=\hspace{-3mm}\sum_{k\in \mathcal{N}[i]}\hspace{-2mm}(Y_{ik}\sum_{n=0}^{\infty} {c_n}^{[k]}s^n) \hspace{2.5mm} \forall i\in  \mathcal{N}-\{r\}\\
\label{conv}
& \displaystyle (\sum_{n=0}^{\infty} c_n^{[i]}s^n)(\sum_{n=0}^{\infty} d_n^{[i]}s^n)=1 \hspace{18mm} \forall i\in  \mathcal{N}-\{r\}
\end{align}
\end{subequations}

The procedure to obtain the coefficients of~\eqref{HELM1} starts by setting $s=0$ in~\eqref{main}. This gives the linear system of $ \sum_{k\in \mathcal{N}[i]}Y_{ik}{c_0}^{[k]}=0$ which always yields the trivial solution (i.e. $c_0^{[i]}=V_{r} \hspace{2mm} \forall i\in  \mathcal{N}-\{r\}$). Next $d_0^{[i]}=1/c_0^{[i]}$ by setting $s=0$ in~\eqref{conv}. The higher order coefficients are progressively obtained by solving the linear system of~\eqref{ln} $(\forall i\in  \mathcal{N}-\{r\})$ which itself is obtained by differentiating~\eqref{main} with respect to $s$ and evaluating at $s=0$ and the convolution formula of~\eqref{conv2}.\vspace{-3mm}
\begin{subequations}\label{HELM2}
\begin{align}
\label{ln}
&S^*_i{d^*_{n-1}}^{[i]}=\sum_{k\in \mathcal{N}[i]}Y_{ik}{c_n}^{[k]}  \hspace{15mm} \forall i\in  \mathcal{N}-\{r\}\\
\label{conv2}
&\displaystyle d_n^{[i]}=\frac{-\sum_{m=0}^{n-1}c_{n-m}^{[i]}d_m^{[i]}}{c_0^{[i]}} \hspace{18.5mm} \forall i\in  \mathcal{N}-\{r\}
\end{align}
\end{subequations}

The radius of convergence is $R=\lim_{n\rightarrow\infty}|c_n|/|c_{n+1}|$, if the limit exists. This marks the distance from the origin to the closest branch point. Notice that when an analytic function does not have a closed-form expression as in this case, its representation as a power series expansion can be approximated by a partial sum of a finite order. Since this approximation for $V_i(s)$ does not converge for $|s|\ge R$ the analytic continuation of these complex functions toward $s=1$ requires an alternative representation of these analytic functions. One such representation, with superior convergence properties, is a continued fraction ($C$-fraction) which is approximated by truncation. The relation between these two representations is crucial for understanding of Pad\'{e} approximants and is described below~\cite{Baker},


For a given power series $V(s)=c_0+c_1s+c_2s^2+...$, assume the existence of the reciprocal relation between the original series, as modified below, and a new series indexed by superscript $^{(1)}$,\vspace{-1mm}
\begin{align}\label{}
\displaystyle 1+\frac{c_2s}{c_1}+\frac{c_3s^2}{c_1}+...= (1+c_1^{(1)}s+c_2^{(1)}s^2+...)^{-1}
\end{align}

Now the original power series can be expressed as,
\vspace{-1mm}
\begin{align}\label{cf1}
\displaystyle c_0+c_1s+c_2s^2+...=c_0+ \displaystyle \frac{c_1s}{1+c_1^{(1)}s+c_2^{(1)}s^2+...}
\end{align}

Next assume the existence of another reciprocal relation between the modified series from the denominator of the fraction in~\eqref{cf1} and a new series indexed by superscript $^{(2)}$,
\vspace{-1mm}
\begin{align}\label{}
\displaystyle1+\frac{c_2^{(1)}s}{c_1^{(1)}}+\frac{c_3^{(1)}s^2}{c_1^{(1)}}+...= \displaystyle (1+c_1^{(2)}s+c_2^{(2)}s^2+...)^{-1}
\end{align}

This allows the expansion of the denominator of~\eqref{cf1} in terms of another fraction,\vspace{-1mm}
\begin{align}\label{}
\displaystyle c_0+c_1s+c_2s^2+...= c_0+  \displaystyle\frac{c_1s}{1+ \displaystyle \frac{c_1^{(1)}s}{1+c_1^{(2)}s+c_2^{(2)}s^2+...}}
\end{align}

By successively forming the reciprocal series we obtain a $C$-fraction, written in a compact form as,\vspace{-3mm}

\begin{align}\label{cfc}
\displaystyle c_0+c_1s+c_2s^2+...=c_0+{\genfrac{}{}{}{}{c_1s}{1}}   {\genfrac{}{}{0pt}{}{}{+}}   {\genfrac{}{}{}{}{c_1^{(1)}s}{1}}   {\genfrac{}{}{0pt}{}{}{+}}   {\genfrac{}{}{}{}{c_1^{(2)}s}{1}}   {\genfrac{}{}{0pt}{}{}{+\dots}}
\end{align}

By truncating the $C$-fraction in~\eqref{cfc} we obtain its convergents which are rational fractions in $s$. For example the first 4 convergents of~\eqref{cfc} are given as,\vspace{-1mm}
\begin{align}\label{}
\nonumber
&\displaystyle \frac{A_0(s)}{B_0(s)}=c_0, \quad \displaystyle \frac{A_1(s)}{B_1(s)}=c_0+c_1s,\\
&\displaystyle \frac{A_2(s)}{B_2(s)}=\displaystyle \frac{c_0+ (c_0c_1^{(1)}+c_1)s}{1+c_1^{(1)}s}, \\
&\displaystyle \frac{A_3(s)}{B_3(s)}=\displaystyle \frac{c_0+ (c_0(c_1^{(1)}+c_1^{(2)})+c_1)s+ c_1c_1^{(2)}s^2}{1+(c_1^{(1)}+c_1^{(2)})s} \nonumber
\end{align}

\noindent
where $c_1^{(1)}=-c_2/c_1$ and $c_1^{(2)}=(c_2^2-c_1c_3)/(c_1c_2)$.


The diagonal Pad\'{e} approximant of degree $M$ of $V(s)$, hereafter appearing frequently in the text, is the (2$M$+1)th convergent of its $C$-fraction representation in~\eqref{cfc},
\vspace{-1mm}
\begin{align}\label{}
\displaystyle \text{PA}[M/M]_V(s)=\frac{A_{2M}(s)}{B_{2M}(s)}
\end{align}

In general a given analytic function can be approximated by PA$[L/M](s)$ where $L$ and $M$ are not necessarily equal,
\vspace{-1mm}
\begin{align}\label{Pade}
\sum_{n=0}^{L+M} c_ns^n=\frac{a_0+a_1s^1+...+a_Ls^L}{b_0+b_1s^1+...+b_Ms^M}+\mathcal{O}(s^{L+M+1}), \hspace{1mm} s \sim 0
\end{align}

Setting $b_0=1$, the denominator coefficients $b_1,...,b_M$ are obtained by cross-multiplying~\eqref{Pade}, equating the coefficients of $s^{L+1}$,$s^{L+2}$,...,$s^{L+M}$ to zero and solving the resulting linear system. Next the numerator coefficients $a_0,a_1,..a_L$ are obtained similarly by equating the coefficients of $s^0$,$s^{1}$,...,$s^{L}$.

\begin{figure}[h!]
    \centering

       \includegraphics[scale=1.55,trim=0cm 0.0cm 0cm 0cm,clip]{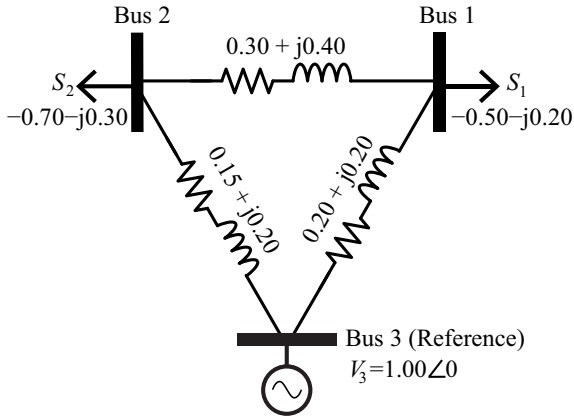}

\caption{3-bus network with no voltage magnitude constraint\vspace{-5mm}} \label{grid1}
\end{figure}

\begin{figure}[h!]
    \centering

       \includegraphics[scale=0.55,trim=0.1cm 0.0cm 0cm 0cm,clip]{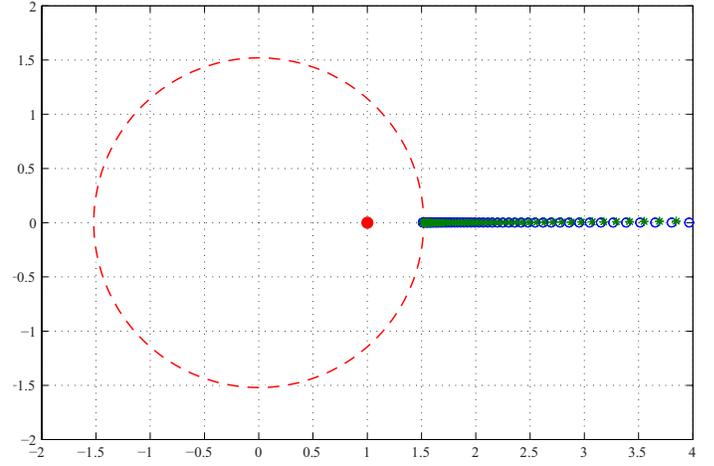}

\caption{Zero-pole distribution of PA[100/100]\vspace{-5mm}} \label{Pade1}
\end{figure}

Now consider the network of Figure~\ref{grid1} where the per-unit values of parameters in~\eqref{complex} are shown. The Taylor series for the unknown states, $V_1(s)$ and $V_2(s)$ are obtained based on~\eqref{HELM2} which are then used to compute the Pad\'{e} coefficients. The concentration of zeros and poles of the diagonal Pad\'{e} approximant, shown in Figure~\ref{Pade1}, defines the closest common branch point of $V_1(s)$ and $V_2(s)$ at $s_\text{b}\hspace{-1mm}=\hspace{-1mm}1.5$ which is also given by Fabry's theorem~\cite{Suetin} as $s_\text{b}\hspace{-1mm}=\hspace{-1mm}\lim_{n\rightarrow\infty}c_n/c_{n+1}$. Here, as it is often the case for the class of problems in~\eqref{complex}, analytic continuation by rational approximation is unnecessary as the power series already converge at $s\hspace{-1mm}=\hspace{-1mm}1$ and thus are sufficient for obtaining $V_1$ and $V_2$. However, PA is a more efficient method as it converges to a given function at a much higher rate than the original power series does~\cite{Suetin}. It can also discover the analytic structure of a given multi-valued function~\cite{Aptekarev}.\vspace{-3mm}

\section{Embedding the Voltage Magnitude Constraints in the Complex Plane}
For a generator node $i \in \mathcal{G}\subset \mathcal{N}$, the real (active) power output, $P_i=\text{Re}(S_i)$, is fixed whereas the imaginary (reactive) power, $Q_i=\text{Im}(S_i)$, is a free parameter which is adjusted so as to fix the magnitude of the voltage phasor $V_i$ at a given setpoint value $M_i$. Notice here the magnitude of a holomorphic function, $V_i(s)$, is to be held fixed which forces it to be constant by open mapping theorem~\cite{Stein} as the image of $V_i(s)$ in the complex plane is a subset of a circle and thus $V_i$ can no longer be an open map. To resolve this contradiction we define an analytic function $\overline{V}_i(s)=\sum_{n=0}^{\infty} \overline{c}_n^{[i]}s^n $ independently of $V_i(s)$ for $i \in \mathcal{G}$ in such way as  $V_i(s)\overline{V}_i(s)=M_i^2$. Note that $\overline{V}_i(s)\neq (V_i(s^*))^*$ and this distinction is the essential concept behind embedding voltage constraints and allows $V_i(s)$ to adopt the value of $V_{r}$ at $s=0$ with its magnitude approaching $M_i$ as $s$ increases. At $s=1$, $\overline{V}_i(s)=(V_i(s^*))^*$ for $i \in \mathcal{G}$.

Since for generator nodes (PV buses), $Q_i=\text{Im}(S_i)$ is unknown we define $S_i(s)$ and $\overline{S}_i(s)$ as functions of complex variable $s$ where $S_i(s)+\overline{S}_i(s)=2P_i$ and introduce~\eqref{polynomial_G} as a complimentary set of equations to~\eqref{p}. At $s=1$,~\eqref{p} combined with~\eqref{polynomial_G} sufficiently determine the AC power flow relations in a network with load and generation. Notice the different embedding of~\eqref{1} and~\eqref{2}. The former ensures that $V_i(s)$ has the trivial solution $V_r$ at $s=0$. The latter enforces $\overline{V}_i(s)=(V_i(s^*))^*$ at $s=1$. Similarly $\overline{S}_i(s)=\sum_{n=0}^{\infty} \overline{g}_n^{[i]}s^n\neq (S_i(s^*))^*$ for $i \in \mathcal{G}$ except in their analytically continued form at $s=1$.\vspace{-5mm}

\begin{subequations}\label{polynomial_G}
\begin{align}
\label{1}
&\displaystyle \frac{s\overline{S}_i(s)}{(V_i(s^*))^*}=\hspace{-2mm}\sum_{k\in \mathcal{N}[i]} V_k(s) Y_{ik} \hspace{28mm} \forall i\in\mathcal{G}\\
\label{2}
&\displaystyle \frac{S_i(s)}{V_i(s)}= \hspace{-1mm}\sum_{k\in \mathcal{N}[i]\cap\mathcal{G}}\hspace{-3mm} \overline{V}_k(s) Y^*_{ik} + \hspace{-3mm}\sum_{k\in \mathcal{N}[i]-\mathcal{G}} \hspace{-3mm}(V_k(s^*))^* Y^*_{ik}\hspace{1mm}\forall i\in\mathcal{G} \\
&V_i(s)\overline{V}_i(s)=M_i^2 \hspace{40mm}\forall i\in\mathcal{G}\\
&S_i(s)+\overline{S}_i(s)=2P_i \hspace{36mm}\forall i\in\mathcal{G}
\end{align}
\end{subequations}

\begin{figure*}[ht]
        \centering

    \subfloat[Zero-pole concentration forms the Stahl's compact set highlighting the common branch points of $V_1(s)$, $\overline{V}_1(s)$, $S_1(s)$ and $V_2(s)$ ($P_1=2.00$).\label{Pade2}]{                \includegraphics  [scale=0.97,trim=0.0cm 0.0cm 0.0cm 0.0cm,clip]{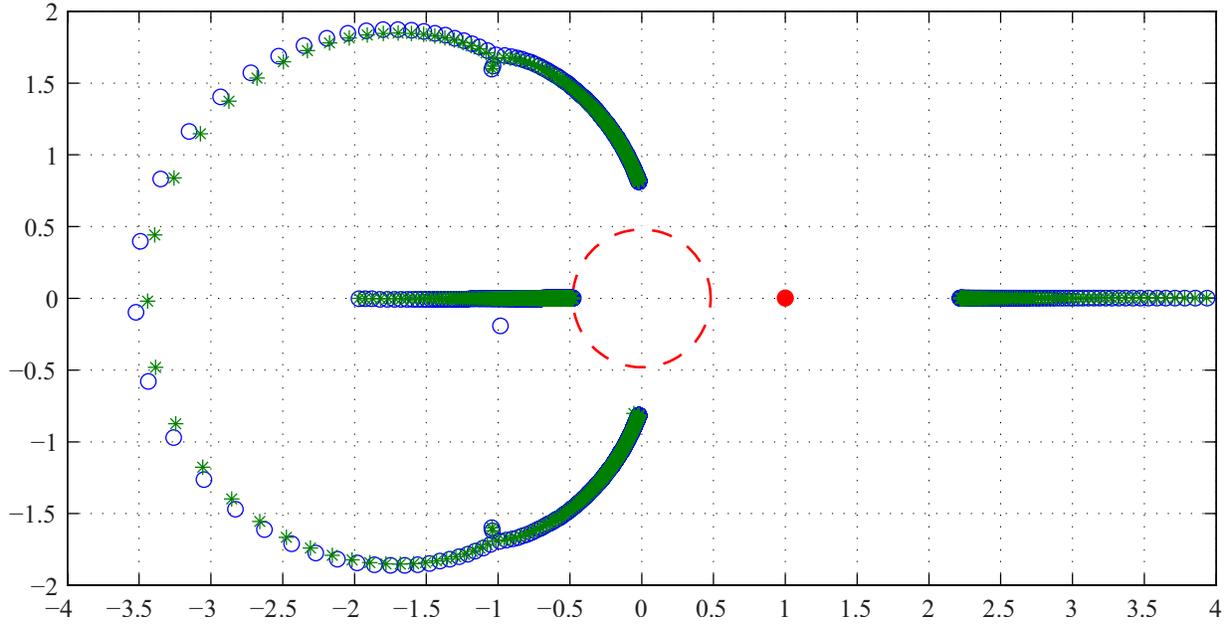}}
           \\
        \subfloat[Transformation of the Stahl's compact set on the feasibility boundary ($P_1=8.36$)\label{Pade_bf}]{
        \includegraphics [scale=0.97,trim=0.0cm 0.cm 0.0cm 0.0cm,clip]{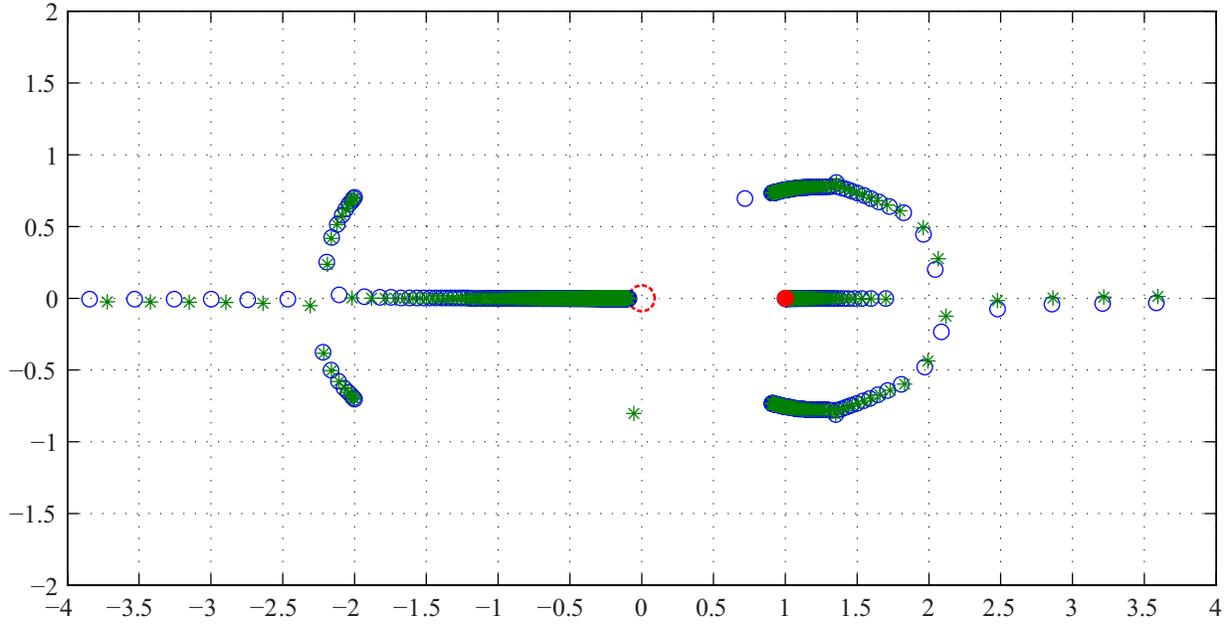}}
         \caption{Zero-pole distribution of PA[1000/1000] depicting the analytic structure of $V_1(s)$  (corresponding to Fig.~\ref{grid2}).\vspace{-2mm}}
\end{figure*}

The combined system of~\eqref{p} and~\eqref{polynomial_G} is adequately described by the following set of power series relations,\vspace{-3mm}

\begin{subequations}\label{HELM1b}
\begin{align}
\label{mainb1}
&sS^*_i\hspace{-1mm}\sum_{n=0}^{\infty}\hspace{-1mm} {d^*_n}^{[i]}s^n\hspace{-0.5mm} =\hspace{-3.0mm}\sum_{k\in \mathcal{N}[i]}\hspace{-2mm}(Y_{ik}\hspace{-1mm}\sum_{n=0}^{\infty}\hspace{-1mm} {c_n}^{[k]}s^n) \hspace{2.5mm} \forall i\in\mathcal{N}\hspace{-1.0mm}-\hspace{-0.50mm}\{r\}\hspace{-0.5mm}-\hspace{-0.5mm}\mathcal{G} \\
\label{mainb2}
&s(\sum_{n=0}^{\infty}\hspace{-0.5mm} \overline{g}_n^{[i]}s^n)(\sum_{n=0}^{\infty}\hspace{-1mm} {d^*_n}^{[i]}s^n)\hspace{-0.5mm} =\hspace{-3.0mm}\sum_{k\in \mathcal{N}[i]}\hspace{-2mm}(Y_{ik}\hspace{-1mm}\sum_{n=0}^{\infty}\hspace{-1mm} {c_n}^{[k]}s^n) \hspace{1.5mm} \forall i\in\mathcal{G}\\
\label{conva}
& \displaystyle (\sum_{n=0}^{\infty} c_n^{[i]}s^n)(\sum_{n=0}^{\infty} d_n^{[i]}s^n)=1 \hspace{17mm} \forall i\in\mathcal{N}-\{r\}\\
\label{convb}
&\displaystyle (\sum_{n=0}^{\infty} c_n^{[i]}s^n)(\sum_{n=0}^{\infty} \overline{c}_n^{[i]}s^n)=M_i^2 \hspace{24.5mm} \forall i\in\mathcal{G}\\
\label{convc}
&(\sum_{n=0}^{\infty} g_n^{[i]}s^n)(\sum_{n=0}^{\infty} d_n^{[i]}s^n) =\nonumber\\
&\hspace{-0.5mm}\sum_{k\in \mathcal{N}[i]\cap\mathcal{G}}\hspace{-3mm}(Y^*_{ik}\hspace{-0.5mm}\sum_{n=0}^{\infty} \overline{c}_n^{[k]}s^n)+\hspace{-3mm}\sum_{k\in \mathcal{N}[i]-\mathcal{G}}\hspace{-3mm}(Y^*_{ik}\hspace{-0.5mm}\sum_{n=0}^{\infty}\hspace{-0.5mm} {c^*_n}^{[k]}s^n) \hspace{2mm} \forall i\in\mathcal{G}\\
&\displaystyle \sum_{n=0}^{\infty} g_n^{[i]}s^n+\sum_{n=0}^{\infty} \overline{g}_n^{[i]}s^n=2P_i \hspace{25mm} \forall i\in\mathcal{G}
\end{align}
\end{subequations}

\begin{figure}
    \centering

       \includegraphics[scale=1.55,trim=0cm 0.0cm 0cm 0cm,clip]{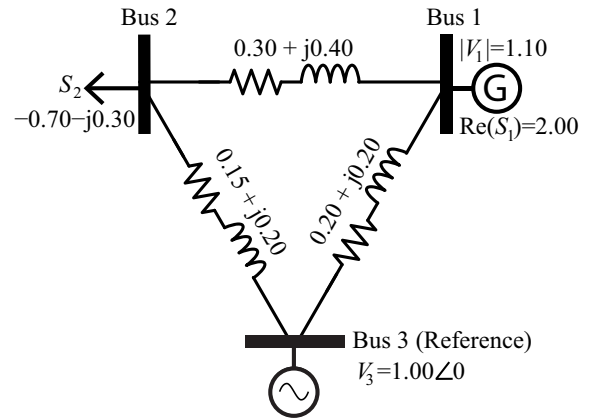}

\caption{3-bus network with voltage control at bus 1\vspace{-5mm}} \label{grid2}
\end{figure}

The key coefficients $c_n^{[i]}$ ($\forall i\in\mathcal{N}\hspace{-1.0mm}-\hspace{-0.50mm}\{r\}$) are progressively obtained by differentiating \eqref{mainb1} and \eqref{mainb2} with respect to $s$, evaluating at $s=0$ and solving the linear system of~\eqref{ln2} and \eqref{ln3} which itself requires the prior knowledge of $d_m^{[i]}$, $\overline{c}_m^{[i]}$, $g_m^{[i]}$ and $\overline{g}_m^{[i]}$ for $m=1,...,n-1$. These coefficients are already obtained at previous stages through~\eqref{conv1b}-\eqref{conv3b}.

Notice that $\overline{g}_0^{[i]}=2P_i-g_0^{[i]}$ whereas for $m\ge1$, $\overline{g}_m^{[i]}=-g_m^{[i]}$.
At $s=0$, we obtain the trivial solution $c_0^{[i]}=V_r$ for all nodes and subsequently $\overline{c}_0^{[i]}=M_i^2/c_0^{[i]}$ for generator nodes.\vspace{-3mm}

\begin{subequations}\label{HELM2b}
\begin{align}
\label{ln2}
&S^*_i{d^*_{n-1}}^{[i]}=\sum_{k\in \mathcal{N}[i]}Y_{ik}{c_n}^{[k]} \hspace{12.5mm} \forall i\in\mathcal{N}\hspace{-1.0mm}-\hspace{-0.50mm}\{r\}\hspace{-0.5mm}-\hspace{-0.5mm}\mathcal{G}\\
\label{ln3}
&\sum_{m=0}^{n-1}\overline{g}_{n-m-1}^{[i]}{d^*}_m^{[i]}=\sum_{k\in \mathcal{N}[i]}Y_{ik}{c_n}^{[k]} \hspace{16mm} \forall i\in\mathcal{G}\\
\label{conv1b}
&\displaystyle d_n^{[i]}=\frac{-\sum_{m=0}^{n-1}c_{n-m}^{[i]}d_m^{[i]}}{c_0^{[i]}} \hspace{20mm} \forall i\in\mathcal{N}-\{r\}\\
\label{conv2b}
&\displaystyle \overline{c}_n^{[i]}=\frac{-\sum_{m=0}^{n-1}c_{n-m}^{[i]}\overline{c}_m^{[i]}}{c_0^{[i]}}  \hspace{31mm} \forall i\in\mathcal{G}\\
\label{conv3b}
&g_{n}^{[i]}=\hspace{-1mm}c_0^{[i]}( \hspace{-3mm}\sum_{k\in \mathcal{N}[i]\cap\mathcal{G}}\hspace{-4mm}Y^*_{ik} \overline{c}_n^{[k]}\hspace{-1mm}+\hspace{-4mm}\sum_{k\in \mathcal{N}[i]-\mathcal{G}}\hspace{-4mm}Y^*_{ik} {c^*_n}^{[k]}\hspace{-1mm}-\hspace{-1mm}\sum_{m=0}^{n-1}\hspace{-1mm}g_{n-m-1}^{[i]}d_m^{[i]})  \hspace{1mm} \forall i\in\mathcal{G}
\end{align}
\end{subequations}

Now consider the modified network of Figure~\ref{grid2} where bus 1 has a generator that regulates its voltage magnitude at $1.10$ and generates $P_1=2.00$. Figure~\ref{Pade2} shows the zero-pole distribution of the diagonal Pad\'{e} approximant for $V_1(s)$ forming the Stahl's compact set~\cite{Baker}. The concentration of zeros and poles highlights the branch points of $V_1(s)$ which, in this case, are common with $\overline{V}_1(s)$, $V_2(s)$ and $S_1(s)$. From each branch point an analytic arc emanates and culminates in a different branch point or in a Chebotarev's point of the Stahl's compact set. The region of convergence is a disk bounded by the closest branch point $s_\text{b}\approx-0.50$. In contrast to the previous case, here, the concept of analytic continuation by Pad\'{e} approximants is elegantly illustrated. Since $\lim_{n\rightarrow\infty}|c_{n+1}|/|c_{n}|\approx 2$, the coefficients tend to explode rapidly.  Without Pad\'{e} approximants based on these otherwise useless coefficients, it is impossible to recover the network solution.

Figure~\ref{Pade_bf} shows the transformation of the Stahl's compact set as $P_1$ reaches the feasibility boundary. The branch point on the positive real axis has now moved to $s=1$. Since past the branch point, there is a non-trivial monodromy, examining the PA solutions, as the degree of the diagonal Pad\'{e} approximants is increased, reveals whether the power flow problem has a stable/high voltage solution or not. The location of this branch point can also serve as a proximity index to the feasibility boundary where the saddle-node bifurcation, i.e. loss of structural stability, occurs. 

\section{Superiority of PA over Newton-Raphson and Semidefinite Relaxation Methods}

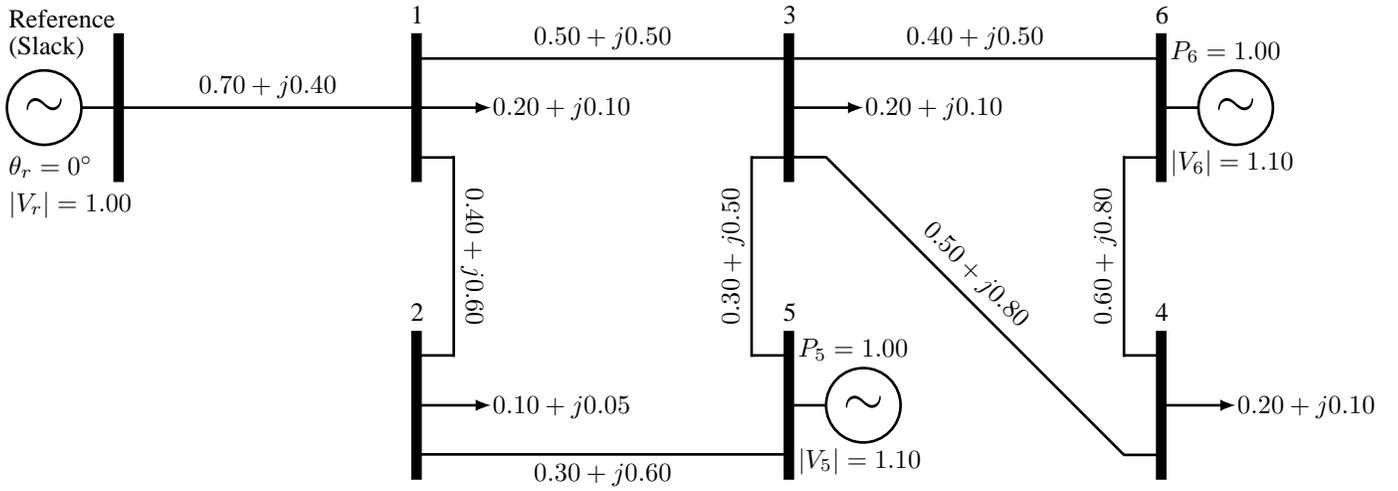
\begin{figure*}
\centering
\begin{circuitikz}[scale=0.99, transform shape]

\path[draw,line width=4pt] (0,0) -- (0,2);
\draw (-1.6,2.2) node[right] {\text{Reference}};
\draw (-1.6,1.8) node[right] {\text{(Slack)}};
\draw (-1.6,-0.3) node[right] {$|V_r| = 1.00$};
\draw (-1.6,0.2) node[right] {$\theta_r = 0^\circ$};
\path[draw,line width=1pt] (-0.5,1) -- (0,1);
\draw[line width=1] (-1,1) circle (0.5);
\draw (-1,1) node{\scalebox{2}{$\sim$}};

\path[draw,line width=4pt] (4,0) -- (4,2);
\draw (3.8,2.25) node[right] {1};
\draw[-latex,line width=1pt] (4,1)-- (5,1);
\draw (4.90,0.98) node[right] {$0.20+j0.10$};

\path[draw,line width=4pt] (9,0) -- (9,2);
\draw (8.8,2.25) node[right] {3};
\draw[-latex,line width=1pt] (9,1)-- (10,1);
\draw (9.9,0.98) node[right] {$0.20+j0.10$};

\path[draw,line width=4pt] (14,0) -- (14,2);
\draw (13.8,2.25) node[right] {6};
\path[draw,line width=1pt] (14.5,1) -- (14,1);
\draw[line width=1] (15,1) circle (0.5);
\draw (15,1) node{\scalebox{2}{$\sim$}};
\draw (14,0.25) node[right] {$|V_6| = 1.10$};
\draw (14,1.75) node[right] {$P_6 = 1.00$};

\path[draw,line width=4pt] (4,-4) -- (4,-2);
\draw (3.8,-1.75) node[right] {2};
\draw[-latex,line width=1pt] (4,-3)-- (5,-3);
\draw (4.9,-3.02) node[right] {$0.10+j0.05$};

\path[draw,line width=4pt] (9,-4) -- (9,-2);
\draw (8.8,-1.75) node[right] {5};
\path[draw,line width=1pt] (9,-3) -- (9.5,-3);
\draw[line width=1] (10,-3) circle (0.5);
\draw (10,-3) node{\scalebox{2}{$\sim$}};
\draw (9,-3.75) node[right] {$|V_5| = 1.10$};
\draw (9,-2.25) node[right] {$P_5 = 1.00$};

\path[draw,line width=4pt] (14,-4) -- (14,-2);
\draw (13.8,-1.75) node[right] {4};
\draw[-latex,line width=1pt] (14,-3)-- (15,-3);
\draw (14.9,-3.02) node[right] {$0.20+j0.10$};
%

\path[draw,line width=1pt] (0,1)-- (4,1);
\draw (2,1.55) node[below] {$0.70+j0.40$};

\path[draw,line width=1pt] (4,1.66)-- (9,1.66);
\draw (6.5,2.20) node[below] {$0.50+j0.50$};

\path[draw,line width=1pt] (9,1.66)-- (14,1.66);
\draw (11.5,2.20) node[below] {$0.40+j0.50$};

\path[draw,line width=1pt] (4,0.33) -- (4.5,0.33);
\path[draw,line width=1pt] (4.5,0.35) -- (4.5,-2.35);
\path[draw,line width=1pt] (4.5,-2.33) -- (4,-2.33);
\draw (4.75,-1) node[rotate=-90] {$0.40+j0.60$};

\path[draw,line width=1pt] (9,0.33) -- (8.5,0.33);
\path[draw,line width=1pt] (8.5,0.35) -- (8.5,-2.35);
\path[draw,line width=1pt] (8.5,-2.33) -- (9,-2.33);
\draw (8.25,-1) node[rotate=90] {$0.30+j0.50$};

\path[draw,line width=1pt] (4,-3.66)-- (9,-3.66);
\draw (6.5,-3.66) node[below] {$0.30+j0.60$};

\path[draw,line width=1pt] (14,0.33) -- (13.5,0.33);
\path[draw,line width=1pt] (13.5,0.35) -- (13.5,-2.35);
\path[draw,line width=1pt] (13.5,-2.33) -- (14,-2.33);
\draw (13.25,-1) node[rotate=90] {$0.60+j0.80$};

\path[draw,line width=1pt] (9,0.334) -- (9.51,0.334);
\path[draw,line width=1pt] (9.5,0.334) -- (13.5,-3.667);
\path[draw,line width=1pt] (13.49,-3.667) -- (14,-3.667);
\draw (11.5,-1.25) node[rotate=-45] {$0.50+j0.80$};

\end{circuitikz}
\caption{7-bus network\vspace{-2mm}}
\label{7b}

\end{figure*}

\begin{table}
\caption{$P_6=0.20$ (all methods finding the stable solution)} \label{t1}
\vspace{-1mm}
\centering
\renewcommand{\arraystretch}{1.1}
\begin{tabular} {|c||c|c|c|c|} \hline 
 Voltage        & Pad\'{e}      & Newton & SDP  & SDP \\
 Magnitude      & Approx.      & Raphson & (1st order)  & (2nd order) \\\hline
$|V_1|$&0.9408  &0.9408 &0.9408 &0.9408 	\\\hline
$|V_2|$&0.9774  &0.9774 &0.9774 &0.9774	\\\hline
$|V_3|$&0.9953  &0.9953 &0.9953 &0.9953	\\\hline
$|V_4|$&0.9447  &0.9447 &0.9447 &0.9447 	\\\hline
\end{tabular}
\vspace{3mm}

\caption{$P_6=0.30$ (failure of first-order relaxation)} \label{t2}
\vspace{-1mm}
\centering
\renewcommand{\arraystretch}{1.1}
\begin{tabular} {|c||c|c|c|c|} \hline 
 Voltage        & Pad\'{e}      & Newton & SDP  & SDP \\
 Magnitude      & Approx.      & Raphson & (1st order)  & (2nd order) \\\hline
$|V_1|$&0.9217  &0.9217&-&0.9217	\\\hline
$|V_2|$&0.9640  &0.9640&-&0.9640	\\\hline
$|V_3|$&0.9897  &0.9897&-&0.9897	\\\hline
$|V_4|$&0.9403  &0.9403 &-&0.9403 	\\\hline
\end{tabular}
\vspace{3mm}

\caption{$P_6=0.75$ (false solution of moment relaxation)} \label{t3}
\vspace{-1mm}
\centering
\renewcommand{\arraystretch}{1.1}
\begin{tabular} {|c||c|c|c|c|} \hline 
 Voltage        & Pad\'{e}      & Newton & SDP  & SDP \\
Magnitude      & Approx.      & Raphson & (1st order)  & (2nd order) \\\hline
$|V_1|$&0.7613  &0.7613&-&0.8504	\\\hline
$|V_2|$&0.8658  &0.8658&-&0.9456	\\\hline
$|V_3|$&0.9210  &0.9210&-&0.7960	\\\hline
$|V_4|$&0.8888  &0.8888 &-&0.1321 	\\\hline
\end{tabular}
\vspace{3mm}

\caption{$P_6=1.00$ (non-convergence of Newton-Raphson) } \label{t4}
\vspace{-1mm}
\centering
\renewcommand{\arraystretch}{1.1}
\begin{tabular} {|c||c|c|c|c|} \hline 
 Voltage        & Pad\'{e}      & Newton & SDP  & SDP \\
 Magnitude      & Approx.      & Raphson & (1st order)  & (2nd order) \\\hline
$|V_1|$&0.5657  &-&-&0.8609	\\\hline
$|V_2|$&0.7546  &-&-&0.9405	\\\hline
$|V_3|$&0.8394  &-&-&0.8178	\\\hline
$|V_4|$&0.8319  &-&-&0.1294	\\\hline
\end{tabular}
\vspace{3mm}

\caption{$P_6=1.02$ (false solution of Newton-Raphson) } \label{t5}
\vspace{-1mm}
\centering
\renewcommand{\arraystretch}{1.1}
\begin{tabular} {|c||c|c|c|c|} \hline 
 Voltage        & Pad\'{e}      & Newton & SDP  & SDP \\
 Magnitude      & Approx.      & Raphson & (1st order)  & (2nd order) \\\hline
$|V_1|$&0.5355  &0.1520&-&0.8575	\\\hline
$|V_2|$&0.7380  &0.0673&-&0.9380	\\\hline
$|V_3|$&0.8283  &0.7376&-&0.8167	\\\hline
$|V_4|$&0.8247  &0.7757&-&0.1296	\\\hline
\end{tabular}
\vspace{3mm}

\caption{$P_6=1.12$ (non-existence of a physical solution)} \label{t6}
\vspace{-1mm}
\centering
\renewcommand{\arraystretch}{1.1}
\begin{tabular} {|c||c|c|c|c|} \hline 
 Voltage        & Pad\'{e}      & Newton & SDP  & SDP \\
 Magnitude     & Approx.      & Raphson & (1st order)  & (2nd order) \\\hline
$|V_1|$&-&0.1242&-&0.8363	\\\hline
$|V_2|$&-&0.0680&-&0.9234	\\\hline
$|V_3|$&-&0.7224&-&0.8086	\\\hline
$|V_4|$&-&0.7609&-&0.1308	\\\hline
\end{tabular}
\vspace{-5mm}
\end{table}

\begin{figure*}
        \centering
        \includegraphics [scale=1.1,trim=0.0cm 0.cm 0.0cm 0.0cm,clip]{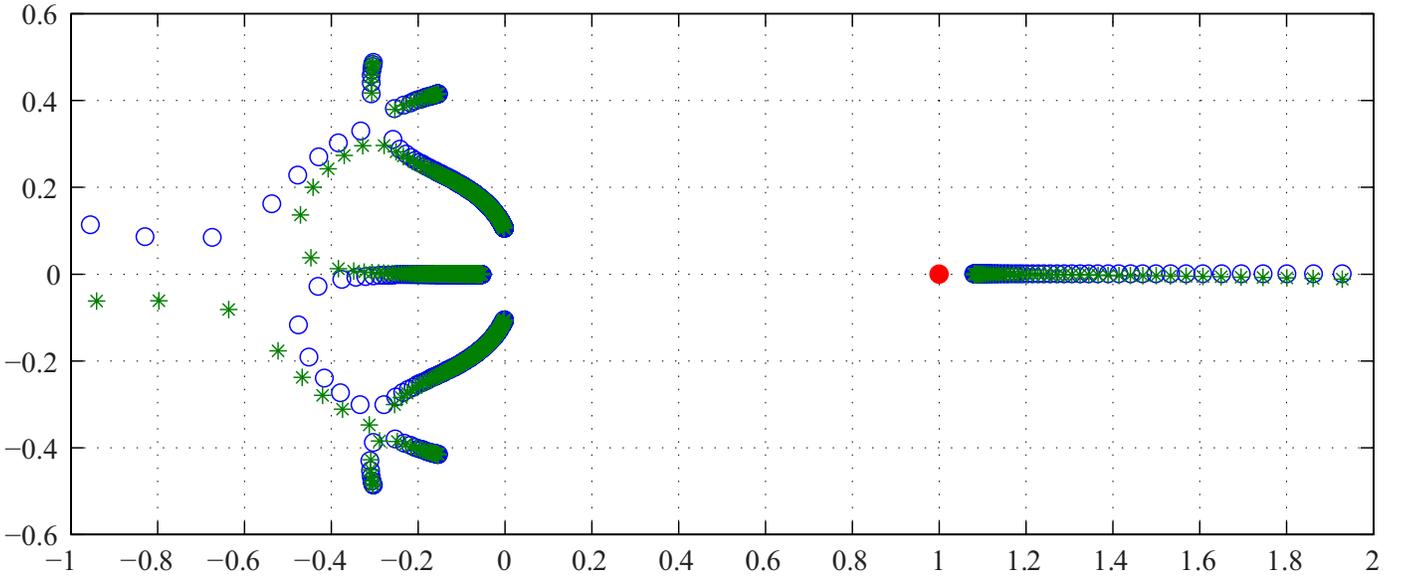}
         \caption{Zero-pole distribution of PA[1000/1000] depicting the analytic structure of $V_6(s)$  (corresponding to bus 6 in Fig.~\ref{7b}).\vspace{2mm}}\label{PA-100p6}
\end{figure*}

Figure~\ref{7b} shows a 7-bus network with 4 load (PQ) buses labeled 1-4, two generator (PV) buses labeled 5 and 6 and a reference (slack) bus. All values are in per unit. Line and load parameters are indicated as complex quantities. The generator voltage magnitudes are controlled at 1.10 and their active power output is 1.00. Newton-Raphson fails to solve this problem as it does not converge with a flat start, i.e. when initialized with all phase angles set to zero and all PQ voltage magnitudes set to 1.00. The first-order semidefinite relaxation also fails as it is not tight enough and the second-order (moment) relaxation finds an unstable/low voltage solution. In contrast PA method finds the desirable solution and, as the zero-pole distribution in Figure~\ref{PA-100p6} clearly demonstrates, the operating point is on the stable branch and still has some margin to power flow infeasibility.

\begin{table*}
\caption{Erratic convergence of Newton-Raphson at $P_6=1.0416...1.0424\approx1.042$} \label{ext}
\centering
\renewcommand{\arraystretch}{1.4}
\begin{tabular} {|c||c|c|c|c|c|c|c|c|c|} \hline
$P_6$  &1.0416&	1.0417&	1.0418&	1.0419&	1.0420&1.0421&	1.0422&	1.0423&	1.0424\\\hline \hline
$|V_1|$&0.3189&	0.4696&	0.2889&	0.1438&	-&0.1438&	0.4887&	0.3210&	0.1436\\\hline
$|V_2|$&0.6256&	0.0520&	0.6478&	0.0675&	-&0.0675&	0.7126&	0.6266&	0.0675\\\hline
$|V_3|$&0.7667&	0.8261&	0.6513&	0.7342&	-&0.7342&	0.8123&	0.7671&	0.7341\\\hline
$|V_4|$&0.7902&	0.8221&	0.1472&	0.7725&	-&0.7725&	0.8147&	0.7904&	0.7724\\\hline
\end{tabular}
\vspace{-3mm}
\end{table*}

\begin{figure*}
        \centering
  \subfloat[Second-order semidefinite (moment) relaxation solutions (green) versus PA solutions (red)]{\label{P6V3a}\includegraphics [scale=0.88,trim=0.0cm 0.cm 0.0cm 0.0cm,clip]{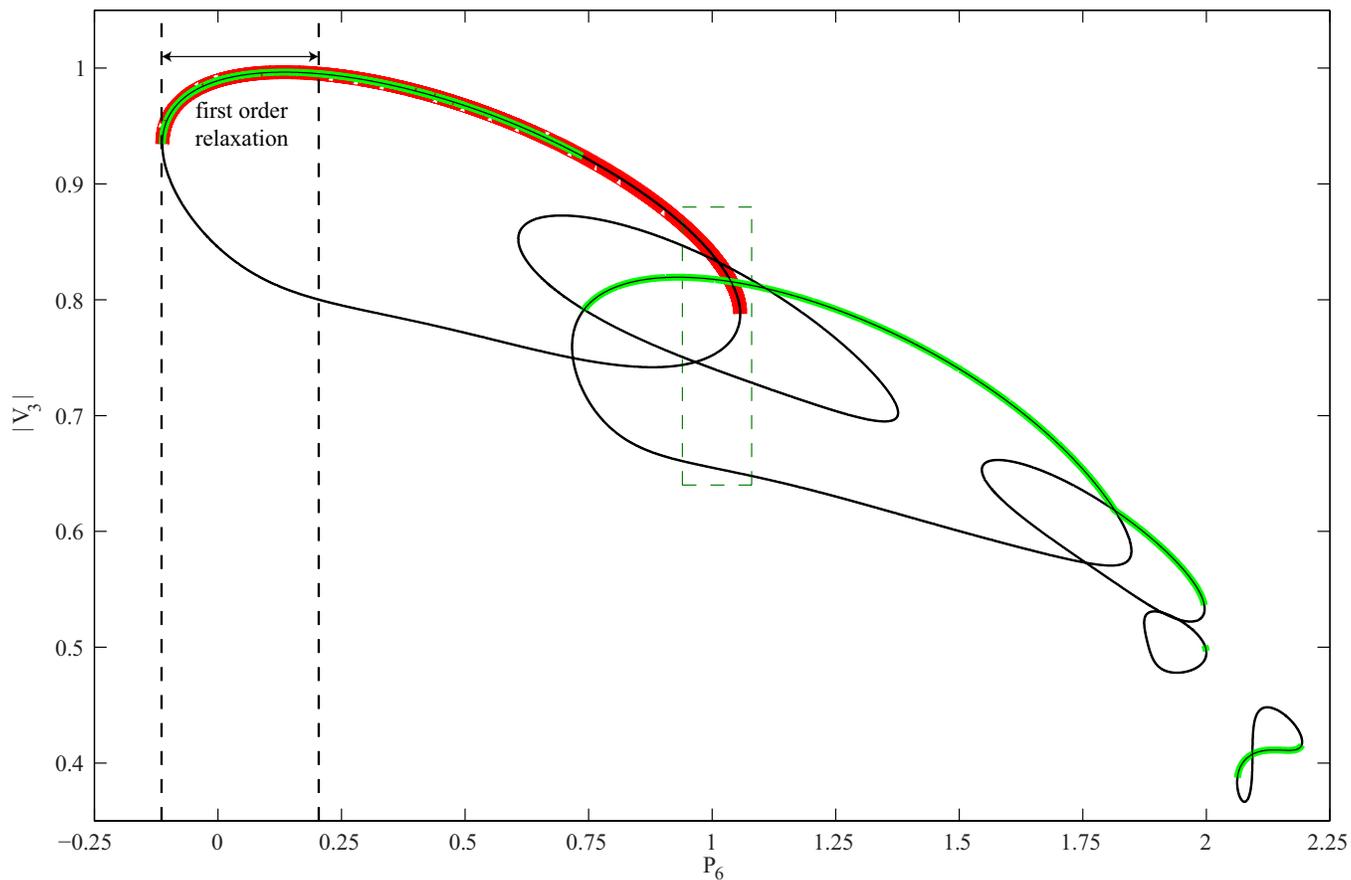}}
                   \\
         \subfloat[Newton-Raphson solutions (blue) versus PA solutions (red)]{\label{P6V3b}\includegraphics  [scale=0.88,trim=0.0cm 0.0cm 0.0cm 0.0cm,clip]{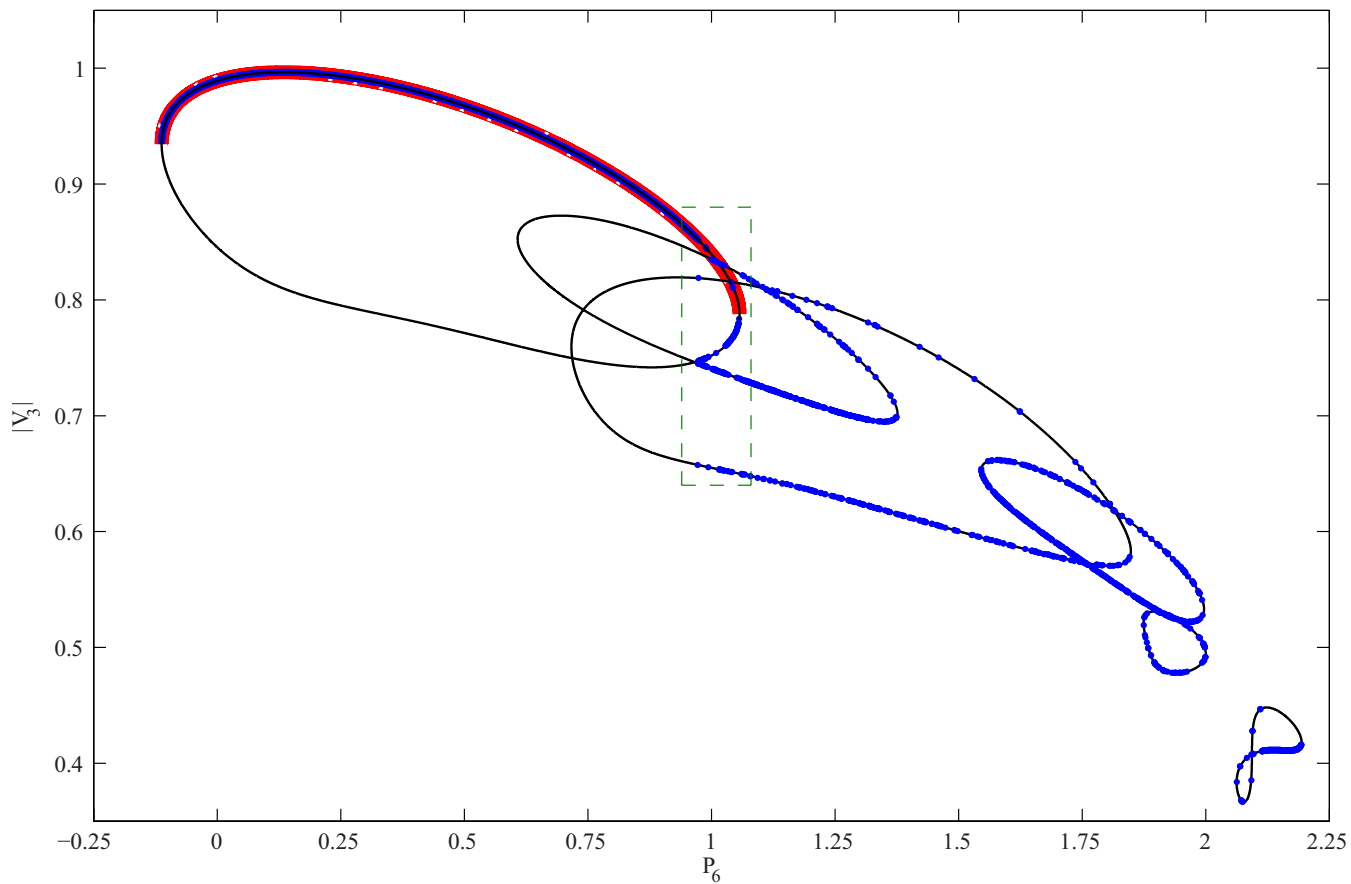}}

         \caption{Superiority of the PA method over Newton-Raphson and SDP methods shown in the context of the network of Fig.~\ref{7b}} \label{P6V3}
\end{figure*}

\begin{figure}
        \centering
           \includegraphics  [scale=0.75,trim=0.10cm 0.0cm 0.0cm 0.0cm,clip]{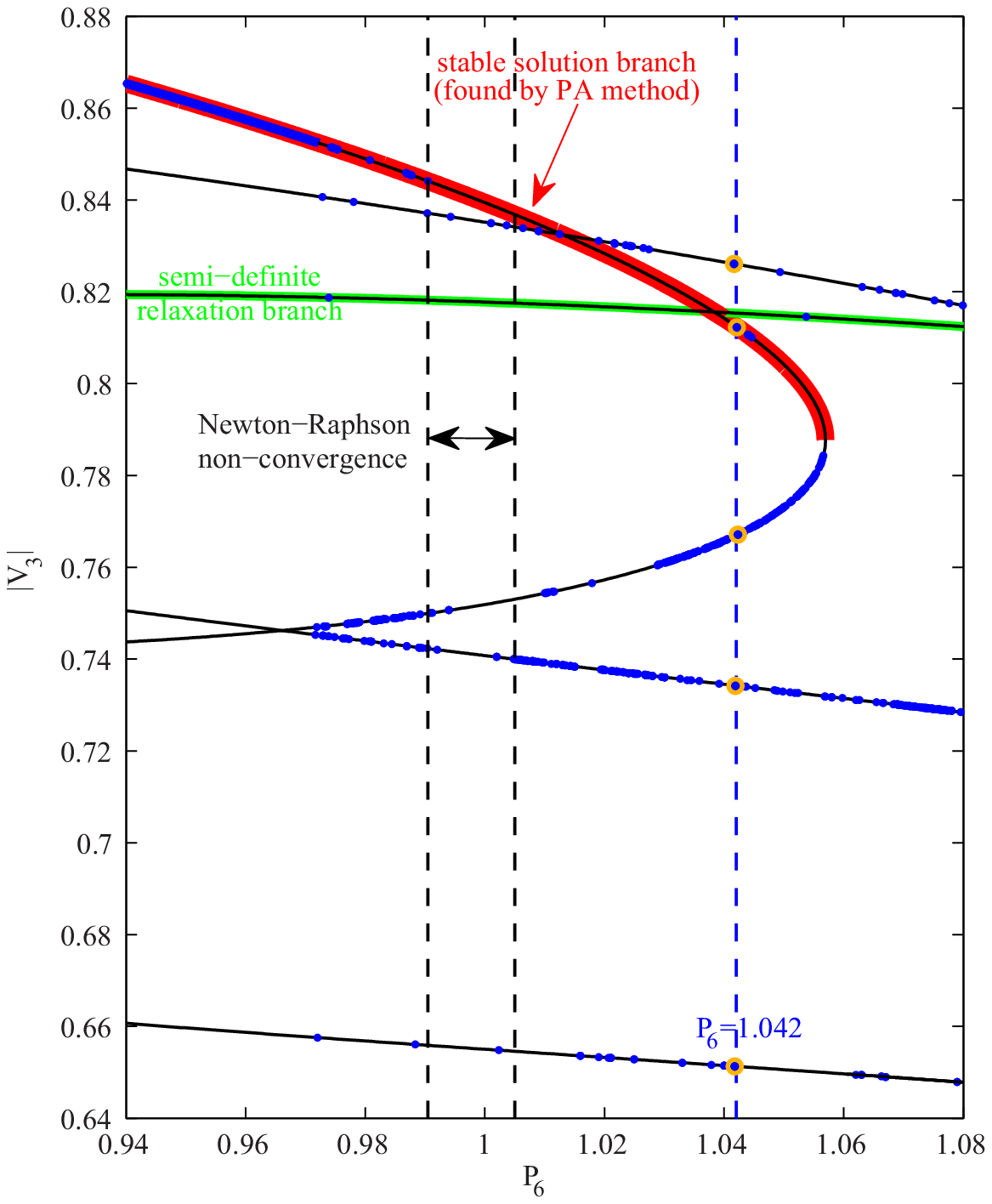}
           \caption{PA solutions (red) versus moment relaxation solutions (green) and Newton-Raphson solutions (blue)\vspace{-4mm}}\label{zoom}
\end{figure}

Now consider the power flow in its polar form~\eqref{polar}. The variables are phase angles of the 6 buses and voltage magnitudes of the 4 load buses. The power flow has 10 equations, relating the active power of the 6 buses and the reactive power of the 4 load buses to phases angles and voltage magnitudes. To better contrast these methods we free a single parameter, $P_6$, the active power generated at bus 6. Each equation defines a hypersurface in $\R^n$ where $n=6+4+1$. The intersection of these hypersurfaces, once projected onto the joint space of the freed parameter and a given variable, yields a series of curves in $\R^2$. Figure~\ref{P6V3} shows these curves in the $(P_6,|V_3|)$ space. The stable/high voltage operating points of the network of Figure~\ref{7b} can only be realized on the segment that is highlighted in red. This segment is consistently found by PA method for all values of $P_6 \in [-0.114 \quad 1.057]$. However Newton-Raphson and semidefinite relaxation methods concurrently find the stable branch only on a small subset of this interval (Table~\ref{t1}). Beyond $P_6=0.204$ the first-order relaxation fails (Table~\ref{t2}). Beyond $P_6=0.739$ the second-order (moment) relaxation finds the false branches (Table~\ref{t3}). These branches are highlighted in green in Figure~\ref{P6V3}a. Newton-Raphson convergence becomes erratic beyond $P_6=0.973$ (Table~\ref{t4}). As Figure~\ref{P6V3}b shows it either does not converge for certain values of $P_6$ (Table~\ref{t4}) or it converges to low-voltage, physically unrealizable and thus false operating points (Table~\ref{t5}). Beyond $P_6=1.057$ there is no physically meaningful solution and the PA method returns no solution whereas both Newton-Raphson and moment relaxation find false solutions (Table~\ref{t6}). It should also be noted that Newton-Raphson is not robust even in finding low-voltage solutions. This is shown in Table~\ref{t6} and highlighted in Figure~\ref{zoom} where small perturbations at $P_6=1.042$ results in Newton-Raphson finding different branches or not converging at all. For industrial applications power flow parameters are typically expressed in 2, 3 and rarely 4 significant digits. Hence the set of values $(1.0416,...,1.0424)$ can be rounded to 4 significant digits and represented as $1.042$ but applying Newton-Raphson to this set yields five topologically distinct solutions as well as non-convergence. Figure~\ref{zoom} shows the detail of the region highlighted by a dashed green box in Figures~\ref{P6V3}a and~\ref{P6V3}b. This region contains the closest bifurcation at $P_6=1.057$ and presents a clear visual contrast between the performance of PA and those of Newton-Raphson and moment relaxation. Notice that Newton-Raphson finds operating points, mostly false, on all 6 solution branches as shown in Figure~\ref{zoom} whereas moment relaxation consistently finds the false branch that Newton-Raphson rarely discovers.

%
%

\section{Conclusion and Ongoing Work}
In this paper we have presented a method of solving the algebraic equations of AC power flow with voltage magnitude constraints. We started by parameterizing the system of equations in the complex plane. This, under the assumption of the algebraic independence of the equations, renders the complex variables, i.e. voltage and power phasors, algebraic in $s$, the complex parameter, and hence expressible as power series in terms of $s$. These power series are in fact Taylor expansions around a trivial solution which defines the germ of the physically realizable and stable solution branch. The radius of convergence of these series is determined by the closest branch point of the algebraic curves that are given as the reduced Groebner basis of the parameterized system of equations. We demonstrated the pivotal role of Pad\'{e} approximants in analytically continuing the germ beyond the region of convergence of the series up to a point of non-trivial monodromy. This enabled the recovery of the physical solution to voltage and power phasors from the divergent power series. Equally important was the analysis of the zero-pole distribution of the Pad\'{e} approximants that reveals the analytic structure of the complex variables. We illustrated how the concentration of zeros and poles form the Stahl's compact set and that the branch point highlighted by the accumulation of zero-poles on the positive real axis serves as a proximity index to power flow infeasibility or voltage collapse. Thus PA method can definitively determine whether a physically realizable and stable solution exists or not.

Next we demonstrated the superiority of the PA method over the conventional Newton-Raphson and the recently developed semidefinite relaxation methods in the context of a 7-bus network. Newton-Raphson, by far the most prevalent method in power industry, can be highly unreliable as the feasibility boundary is approached. It can also converge to solutions that are physically unrealizable and should thus be discarded as false. On the other hand, semidefinite relaxation methods are superior to Newton-Raphson in the sense that they do not exhibit erratic convergence behavior. We are aware of cases where, unlike the 7-bus network examined here, moment relaxation performs substantially better than Newton-Raphson in finding the stable branch. However, in general SDP methods can also fail, i.e. the relaxation is not sufficiently tight, or may consistently find the false branches. While the performance of these optimization methods can potentially be improved by better choices of objective function or adding extra constraints, we argue that the tendency to find false branches is a fundamental limitation of semidefinite relaxation methods which have their origins in real algebraic geometry and therefore cannot utilize the elegant concepts of analytic continuation and monodromy to solve problems of a complex analytic nature.

We are currently investigating the generic structure of the Stahl's compact set for the parameterized algebraic equations of power flow and the effectiveness of PA near the feasibility boundary. We will disseminate our findings in subsequent publications. This will also include a more expanded treatment of network controllers primarily the on-load tap-changing and phase-shifting transformers.
\label{l:conclusion}

\section*{Acknowledgment}
This work is supported in part by the Russian Science Foundation (RSF) under the grant 14-50-00005. The authors would like to thank Nikolay Ikonomov for sharing the code for computing the zeros and poles of Pad\'{e} approximants and Dan Molzahn for sharing the code for solving the power flow problem based on moment relaxation and for reading the manuscript and providing useful comments.


\begin{thebibliography}{4}
\bibitem{Trias}
A. Trias, ``The Holomorphic Embedding Load Flow method," {\em Proceedings of Power and Energy Society General Meeting}, 22-26 July 2012.

\bibitem{Markushevich}
A.I. Markushevich, {\em Theory of Functions of a Complex Variable}, Translated by R.A. Silverman, 2nd Edition, American Mathematical Society, 2005.

\bibitem{Venikov}
V.A. Venikov, V.A. Stroev, V.I. Idelchick, and V.I. Tarasov, ``Estimation of electrical power system steady-state stability in load flow calculations," {\em IEEE Transactions on Power Apparatus and Systems}, vol.94, no.3, pp.1034-1041, May 1975.

\bibitem{Sauer}
P.W. Sauer and M.A. Pai, ``Power system steady-state stability and the load-flow Jacobian," {\em IEEE Transactions on Power Systems}, vol.5, no.4, pp.1374-1383, Nov. 1990.

\bibitem{VC}
I.A. Dobson, et al.,``Chapter 2: Basic Theoretical Concepts," in {\em Voltage Stability Assessment: Concepts, Practices and Tools}, IEEE-PES, 2002.


\bibitem{SB}
S.S. Baghsorkhi, ``Computing Saddle-Node and Limit-Induced Bifurcation Manifolds for Subtransmission and Transmission Wind Generation," \emph{Proceedings of the IEEE Power and Energy Society General Meeting}, Denver, CO, July 2015.

\bibitem{Sauer2}
P.W. Sauer, B.C. Lesieutre and M.A. Pai,  ``Maximum Loadability and Voltage Stability in Power Systems,"  {\em International Journal of Electrical Power and Energy Systems},  vol. 15,  pp.145-154 1993.

\bibitem{Thorp}
J.S. Thorp and S.A. Naqavi, ``Load-flow fractals draw clues to erratic behavior," {\em IEEE Computer Applications in Power}, pp. 59-62, Jan. 1997.

\bibitem{DM}
D.K. Molzahn, ``Application of Semidefinite Optimization Techniques to Problems in Electric Power Systems," Ph.D. Dissertation, University of Wisconsin-Madison, Department of Electrical Engineering, August 2013.

\bibitem{Lasserre}
J.B. Lasserre, {\em Moments, Positive Polynomials and Their Applications}, Imperial College Press, 2010.

\bibitem{Cox} D. Cox, J. Little and D. O'Shea, \emph{Ideals, Varieties, and Algorithms}, 3rd Edition, Springer, 2006.


\bibitem{Baker}
G.A. Baker and P. Graves-Morris, {\em Pad\'{e} Approximants}. Cambridge Univ. Press, 1996.

\bibitem{Stein}
E.M. Stein and R. Shakarchi, {\em Complex Analysis}.  Princeton Univ. Press, 2003.

\bibitem{Suetin}
S. P. Suetin, ``Pad\'{e} approximants and the effective analytic continuation of a power series," \emph{Russian Math. Surveys}, 57:1 (2002), 43-141.

\bibitem{Aptekarev}
A.I. Aptekarev, V.I. Buslaev, A. Martinez-Finkelshtein, S.P. Suetin, ``Pad\'{e} approximants, continued fractions, and orthogonal polynomials," \emph{Russian Math. Surveys}, 66:6 (2011), 1049-1131.

\end{thebibliography}
\end{document}